\documentclass[a4paper,11pt]{article}
\pdfoutput=1 
\usepackage{jcappub} 
\usepackage{lmodern} 
\usepackage[T1]{fontenc}
\usepackage{amsmath}
\usepackage{amssymb}

\title{Relative Likelihood for Life as a Function of Cosmic Time}

\author[a]{Abraham Loeb,}
\author[b]{Rafael A. Batista,}
\author[b]{David Sloan}

\affiliation[a]{Astronomy department, Harvard University, 60 Garden
  Street, Cambridge, MA 02138, USA} 
\affiliation[b]{Department of Physics - Astrophysics, University of
  Oxford,  DWB, Keble Road, OX1  3RH, Oxford, UK}

\emailAdd{aloeb@cfa.harvard.edu}
\emailAdd{rafael.alvesbatista@physics.ox.ac.uk}
\emailAdd{david.sloan@physics.ox.ac.uk}

\abstract{Is life most likely to emerge at the present cosmic time
  near a star like the Sun? We address this question by calculating
  the relative formation probability per unit time of habitable
  Earth-like planets within a fixed comoving volume of the Universe,
  $dP(t)/dt$, starting from the first stars and continuing to the
  distant cosmic future. We conservatively restrict our attention to
  the context of ``life as we know it'' and the standard cosmological
  model, $\Lambda$CDM.  We find that unless habitability around low
  mass stars is suppressed, life is most likely to exist near $\sim
  0.1M_\odot$ stars ten trillion years from now. Spectroscopic
  searches for biosignatures in the atmospheres of transiting Earth-mass
  planets around low mass stars will determine whether present-day
  life is indeed premature or typical from a cosmic perspective.}


\begin{document}

\maketitle

\section{Introduction}
\label{Sec:intro}

The known forms of terrestrial life involve carbon-based chemistry in
liquid water~\cite{Kasting}. In the cosmological context, life could
not have started earlier than 10 Myr after the Big Bang ($z\gtrsim
140$) since the entire Universe was bathed in a thermal radiation
background above the boiling temperature of liquid water. Later on,
however, the Universe cooled to a {\it habitable epoch} at a
comfortable temperature of 273-373 K between 10-17 Myr after the Big
Bang~\cite{Loeb14}.

The phase diagram of water allows it to be liquid only under external
pressure in an atmosphere which can be confined gravitationally on the
surface of a planet. To keep the atmosphere bound against evaporation
requires strong surface gravity of a rocky planet with a mass
comparable to or above that of the Earth~\cite{SchallerBrown}.

Life requires stars for two reasons. Stars are needed to produce the
heavy elements (carbon, oxygen and so on, up to iron) out of which
rocky planets and the molecules of life are made. Stars also provide a
heat source for powering the chemistry of life on the surface of their
planets. Each star is surrounded by a habitable zone where the surface
temperature of a planet allows liquid water to exist. The approximate
distance of the habitable zone, $r_{\rm HZ}$, is obtained by equating
the heating rate per unit area from the stellar luminosity, $L$, to the
cooling rate per unit area at a surface temperature of $T_{\rm HZ}\sim
300$~K, namely $({L/ 4\pi r_{\rm HZ}^2})\sim \sigma T_{\rm HZ}^4$,
where $\sigma$ is the Stefan-Boltzman constant~\cite{Kasting}.

According to the standard model of cosmology, the first stars in the
observable Universe formed $\sim 30$ Myr after the Big Bang at a
redshift, $z\sim 70$~\cite{Loeb14,LF13,Fialkov,Naoz}. Within a few
Myr, the first supernovae dispersed heavy elements into the
surrounding gas, enriching the second generation stars with heavy
elements. Remnants from the second generation of stars are found in
the halo of the Milky Way galaxy, and may have planetary systems in
the habitable zone around them~\cite{MashianLoeb}. The related planets
are likely made of carbon, and water could have been delivered to
their surface by icy comets, in a similar manner to the solar system. The
formation of water is expected to consume most of the oxygen in
the metal poor interstellar medium of the first galaxies~\cite{Bialy}.

Currently, we only know of life on Earth.  The Sun formed $\sim 4.6$
Gyr ago and has a lifetime comparable to the current age of the
Universe. But the lowest mass stars near the hydrogen burning
threshold at $0.08M_\odot$ could live a thousand times longer, up to
10 trillion years~\cite{laughlin1997a,Rushby}.  Given that habitable
planets may have existed in the distant past and might exist in the
distant future, it is natural to ask: {\it what is the relative
  probability for the emergence of life as a function of cosmic time?}
In this paper, we answer this question conservatively by restricting
our attention to the context of ``life as we know it'' and the
standard cosmological model ($\Lambda$CDM).\footnote{We address this
  question from the perspective of an observer in a single comoving
  Hubble volume formed after the end of inflation. As such we do not
  consider issues of self-location in the multiverse, nor of the
  measure on eternally inflating regions of space-time. We note,
  however, that any observers in a post-inflationary bubble will by
  necessity of the eternal inflationary process, only be able to
  determine the age of their own bubble. We therefore restrict our
  attention to the question of the probability distribution of life in
  the history of our own inflationary bubble.} Note that since the
probability distribution is normalized to have a unit integral, it
only compares the relative importance of different epochs for the
emergence of life but does not calibrate the overall likelihood for
life in the Universe. This feature makes our results robust to
uncertainties in normalization constants associated with the
likelihood for life on habitable planets.

In \S \ref{Sec:2} we express the relative likelihood for the
appearance of life as a function of cosmic time in terms of the star
formation history of the Universe, the stellar mass function, the
lifetime of stars as a function of their mass, and the probability of
Earth-mass planets in the habitable zone of these stars. We define
this likelihood within a fixed comoving volume which contains a fixed
number of baryons.  In predicting the future, we rely on an
extrapolation of star formation rate until the current gas reservoir
of galaxies is depleted. Finally, we discuss our numerical results in
\S \ref{Sec:3} and their implications in \S \ref{Sec:4}.

\section{Formalism}
\label{Sec:2}

\subsection{Master Equation}

We wish to calculate the probability $dP(t)/dt$ for life to form on
habitable planets per unit time within a fixed comoving volume of the
Universe. This probability distribution should span the time interval
between the formation time of the first stars ($\sim 30$~Myr after the
Big Bang) and the maximum lifetime of all stars that were ever made
($\sim 10$~Tyr).

The probability $dP(t)/dt$ involves a convolution of the star
formation rate per comoving volume, $\dot{\rho}_*(t')$, with the
temporal window function, $g(t-t',m)$, due to the finite lifetime of
stars of different masses, $m$, and the likelihood, $\eta_{\rm Earth}(m)$,
of forming an Earth-mass rocky planet in the habitable zone (HZ) of
stars of different masses, given the mass distribution of stars,
$\xi(m)$, times the probability, $p({\rm life|HZ})$, of actually having
life on a habitable planet. With all these ingredients, the relative
probability per unit time for life within a fixed comoving volume can
be written in terms of the double integral,
\begin{equation}
	\dfrac{dP}{dt} (t) = \dfrac{1}{N} \int\limits_0^{t} dt'
        \int\limits_{m_{min}}^{m_{max}} dm' \xi(m') \dot{\rho}_*
        (t',m') \eta_{\text{Earth}}(m') p(\text{life} | \text{HZ})
        g(t-t', m'),
	\label{eq:prob}
\end{equation}
where the pre-factor $1/N$ assures that the probability
distribution is normalized to a unit integral over all times.  The
window function, $g(t-t',m)$, determines whether a habitable planet
that formed at time $t'$ is still within a habitable zone at time
$t$. This function is non-zero within the lifetime $\tau_*(m)$ of each
star, namely $g(t-t', m) = 1$ if $0<(t - t') < \tau_*(m)$, and zero
otherwise.  The quantities $m_{min}$ and $m_{max}$ represent the
minimum and maximum masses of viable host stars for habitable planets,
respectively. Below we provide more details on each of the various
components of the above master equation.

\subsection{Stellar Mass Range}

Life requires the existence of liquid water on the surface of
Earth-mass planets during the main stage lifetime of their host
star. These requirements place a lower bound on the lifetime of the
host star and thus an upper bound on its mass.

There are several proxies for the minimum time needed for life to
emerge. Certainly the star must live long enough for the planet to
form, a process which took $\sim 40$ Myr for
Earth~\cite{NatureLife}. Moreover, once the planet formed, sufficient
cooling must follow to allow the condensation of water on the planet's
surface. The recent discovery of the earliest crystals, Zircons,
suggest that these were formed during the Archean era, as much as 160
Myr after the planet formed~\cite{Geology}. Thus, we arrive at a
conservative minimum of 200 Myr before life could form - any star
living less than this time could not host life on an Earth-like
planet. At the other end of the scale, we find that the earliest
evidence for life on Earth comes from around 800 Myr after the
formation of the planet~\cite{NatureLife}, yielding an upper bound on
the minimum lifetime of the host star. For the relevant mass range of
massive stars, the lifetime, $\tau_*$, scales with stellar mass, $m$,
roughly as $(\tau_*/\tau_\odot) = (m/M_\odot)^{-\alpha}$, where 
 $\alpha$ depends on the stellar mass, as shown in equation~\ref{eq:lt}.
Thus, we find that the maximum mass of a
star capable of hosting life ($m_{max}$) is in the range $2.8$--$4.7
M_\odot$. Due to their short lifetimes and low abundances, high mass
stars do not provide a significant contribution to the probability
distribution, $dP(t)/dt$, and so the exact choice of the upper mass
cutoff in the above range is unimportant. The lowest mass stars above
the hydrogen burning threshold have a mass $m=0.08 M_\odot$.

\subsection{Time Range}

The first stars are predicted to have formed at a redshift of $z\sim
70$, about $30$ Myr after the Big
Bang~\cite{Loeb14,LF13,Fialkov,Naoz}. Their supernovae resulted in a
second generation of stars -- enriched by heavy elements, merely a few
Myr later. The theoretical expectation that the second generation
stars should have hosted planetary systems can be tested
observationally by searching for planets around metal poor stars in
the halo of the Milky Way galaxy~\cite{MashianLoeb}.

Star formation is expected to exhaust the cold gas in galaxies as soon
as the Universe will age by a factor of a few (based on the ratio
between the current reservoir of cold gas in galaxies~\cite{Fukugita}
and the current star formation rate), but low mass stars would survive
long after that. The lowest mass stars near the hydrogen burning limit
of $0.08M_\odot$, have a lifetime of order 10 trillion
years~\cite{laughlin1997a}. The probability $dP(t)/dt$ is expected to
vanish beyond that time.

\begin{figure}
	\centering
	\includegraphics[width=.49\textwidth]{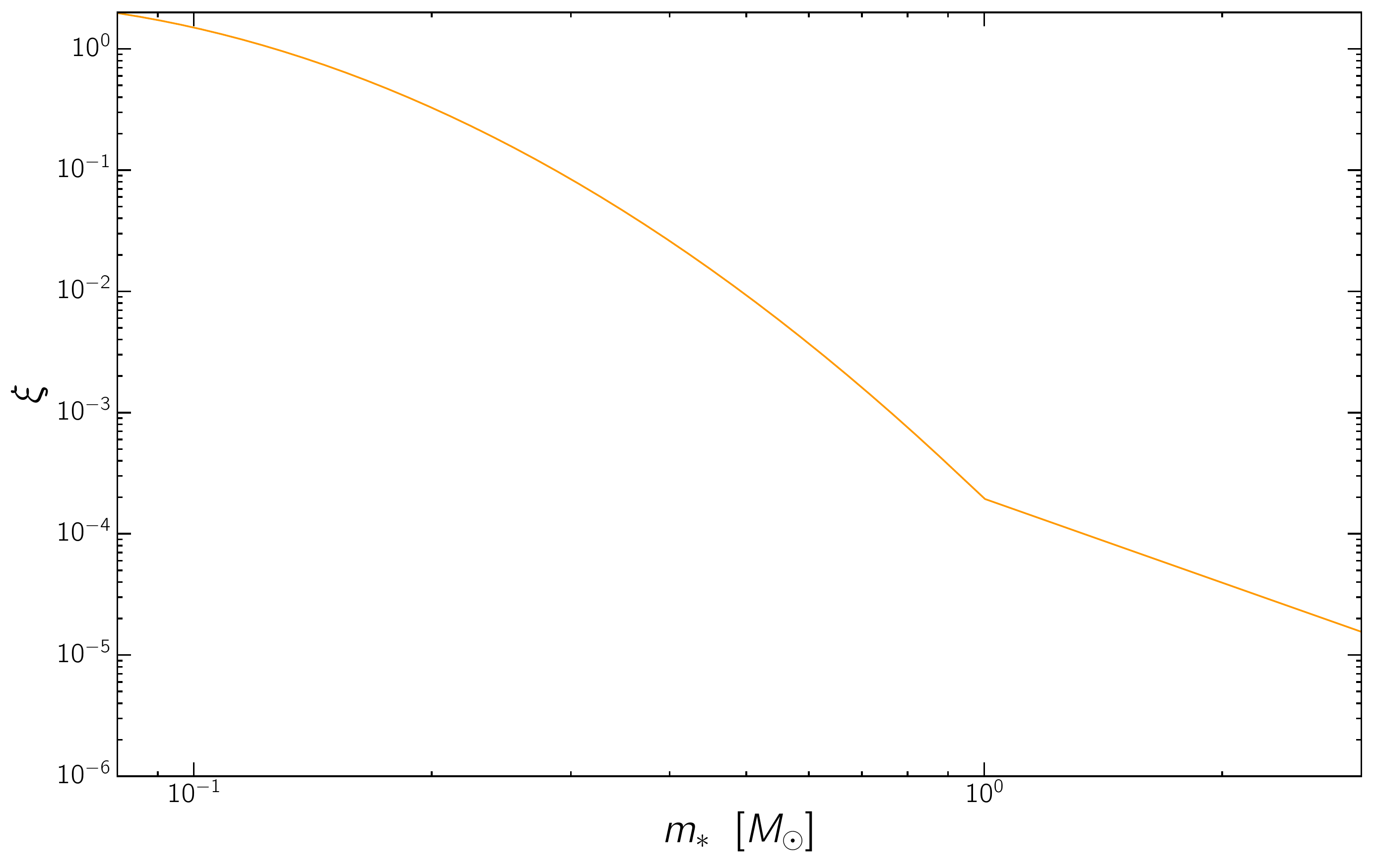}
	\caption{The Chabrier~\cite{Chabrier} mass function of stars,
          $\xi (m_*)$, plotted with a normalization integral of
          unity.}
	\label{fig:imf}
\end{figure}

\subsection{Initial Mass Function}

The initial mass function (IMF) of stars $\xi(m)$ is proportional to
the probability that a star in the mass range between $m$ and $m+dm$
is formed. We adopt the empirically-calibrated, Chabrier functional
form ~\cite{Chabrier}, which follows a lognormal form for masses under
a solar mass, and a power law above a solar mass, as follows:
\begin{equation}
	\xi(m) \propto 
	\begin{cases} 
		\left( \dfrac{m}{M_{\odot}} \right)^{-2.3} \quad\quad\quad\quad\quad\quad m > 1~M_\odot \\
		a \exp\left(-\dfrac{\ln(m/m_c)^2}{2\sigma^2}\right) \dfrac{M_\odot}{m} \quad m \leq  1~M_\odot 
		\end{cases},
\end{equation}
where $a = 790$, $\sigma = 0.69$, and $m_c = 0.08 M_\odot$. This IMF
is plotted as a probability distribution normalized to a unit
integral in Figure~\ref{fig:imf}.

For simplicity, we ignore the evolution of the IMF with cosmic time
and its dependence on galactic environment (e.g., galaxy type or
metallicity~\cite{Conroy}), as well as the uncertain dependence of the
likelihood for habitable planets around these stars on
metallicity~\cite{MashianLoeb}.

\subsection{Stellar Lifetime}

\begin{figure}
	\centering
	\includegraphics[width=.49\textwidth]{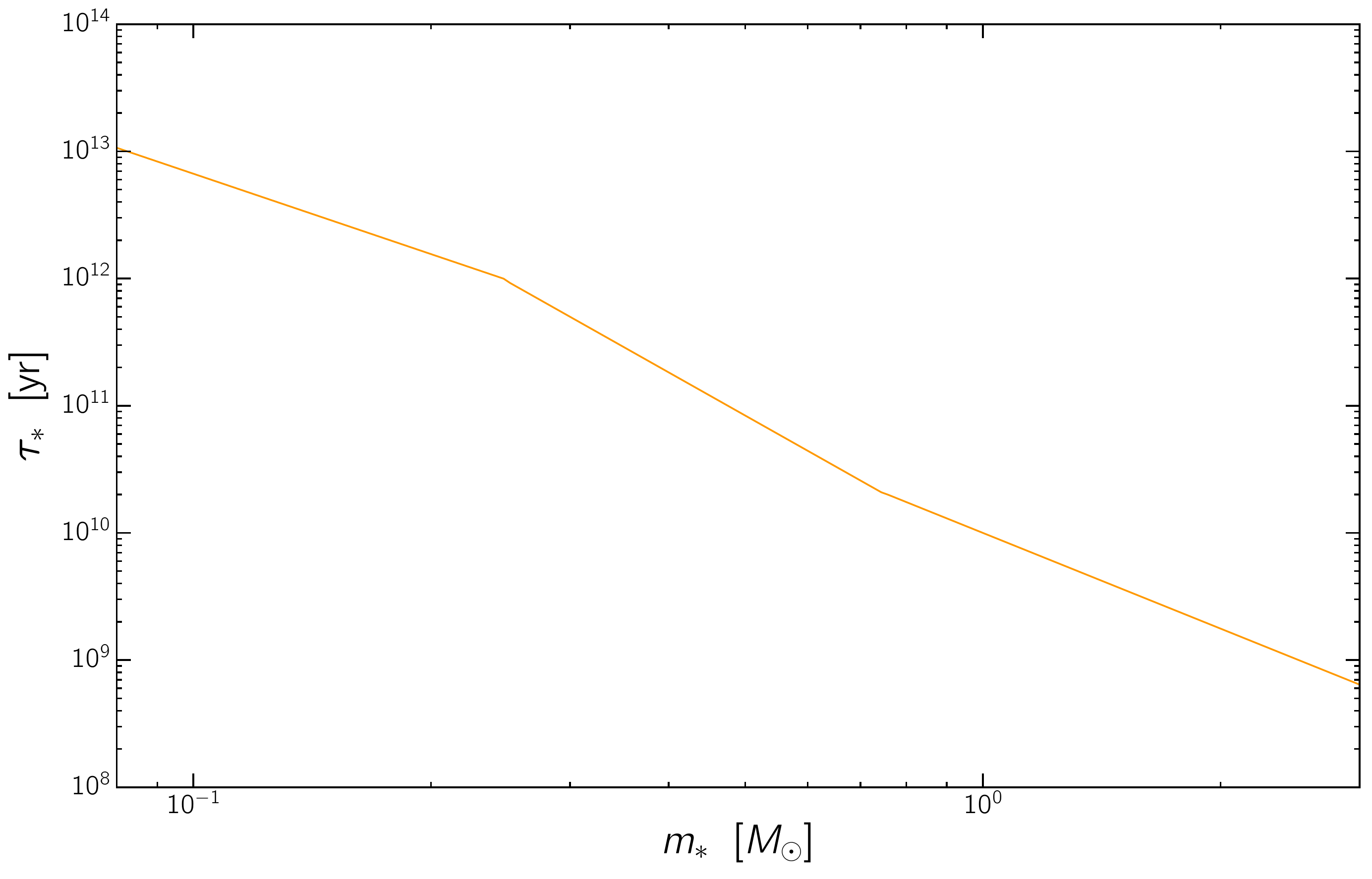}
	\caption{Stellar lifetime ($\tau_*$) as a function of 
          mass.}
	\label{fig:lt}
\end{figure}

The lifetime of stars, $\tau_*$, as a function of their mass, $m$, can
be modelled through a piecewise power-law form. For $m<0.25 M_\odot$,
we follow Ref.~\cite{laughlin1997a}. For $0.75 M_{\odot}<m<3M_\odot$,
we adopt a scaling with an average power law index of -2.5 and the
proper normalization for the Sun~\cite{Salaris}. Finally, we
interpolate in the range between 0.25 and 0.75 $M_\odot$ by fitting a
power-law form there and enforcing continuity. In summary, we adopt,
\begin{equation}
        \tau_*(m)=
	\begin{cases} 
       1.0 \times 10^{10} \; \text{yr} \left(\dfrac{m}{M_\odot}\right)^{-2.5} \quad 0.75 M_\odot < m < 3 M_\odot \\
       7.6 \times 10^{9} \; \text{yr} \left(\dfrac{m}{M_\odot}\right)^{-3.5} \quad 0.25   M_\odot < m \leq 0.75 M_\odot  \\
       5.3 \times 10^{10} \; \text{yr} \left(\dfrac{m}{M_\odot}\right)^{-2.1} \quad 0.08 M_\odot\leq m \leq  0.25 M_\odot 
	\end{cases}	.
	\label{eq:lt}
\end{equation}
This dependence is depicted in Figure~\ref{fig:lt}.

The metallicity of galaxies increases steadily as stars
  fuse hydrogen and helium into heavier elements. A metallicity of
  $\mathcal{Z} \approx 0.04$, about twice the solar value, maximizes
  the lifetime of stars~\cite{adams1997a}. We have incorporated the
  metallicity dependence of stellar lifetime into our calculations and
  found the resulting change in our results to be small compared to other
  uncertainties. For simplicity, we therefore ignore the metallicity
  evolution in our calculations. 

\subsection{Star Formation Rate}

We adopt an empirical fit to the star formation rate per comoving
volume as a function of redshift, $z$~\cite{MadauDickinson},
\begin{equation}
	\dot{\rho}_*(z) = 0.015
        \dfrac{(1+z)^{2.7}}{1+[(1+z)/2.9]^{5.6}}  \; M_\odot
        \text{yr}^{-1} \text{Mpc}^{-3} ,
\end{equation}
and truncate the extrapolation to early times at the expected
formation time of the first stars~\cite{Loeb14}. We extrapolate the
cosmic star formation history to the future or equivalently negative
redshifts $-1 \le z<0$ (see, e.g. Ref.~\cite{barnes2005a}) and find
that the comoving star formation rate drops to less than $10^{-5}$ of
the current rate at 56 Gyr into the future. We cut off the star
formation at roughly the ratio between the current reservoir mass of
cold gas in galaxies~\cite{Fukugita} and the current star formation
rate per comoving volume. 

To avoid an abrupt cutoff in the star formation rate, we
  assume a simple exponential form, $\exp\{-(t/t_1)\}$, with a
  characteristic timescale of $t_1 \sim 50 \; \text{Gyr}$ that is
  dictated by the ratio between the present-day gas reservoir and star
  formation rate. This form is appropriate for a closed box model in
  which the consumption rate of gas in star formation is proportional
  to the gas mass available. The infall of fresh gas into galaxies is
  heavilly suppressed in the cosmic future due to the accelerated
  cosmic expansion~\cite{Nagamine,Busha}. Although galaxies continue
  to consume their gas reservoirs through star formation, some of this
  gas may be lost through supernova or quasar-driven winds. We
  therefore considered also a sharper exponential form, $\exp\{-(t/t_2)\}$
  with $t_2\sim 20~{\rm Gyr}$, but found the final results to be
  indistinguishable within the overall uncertainties of the model.

\begin{figure}
	\centering
	\includegraphics[width=.49\textwidth]{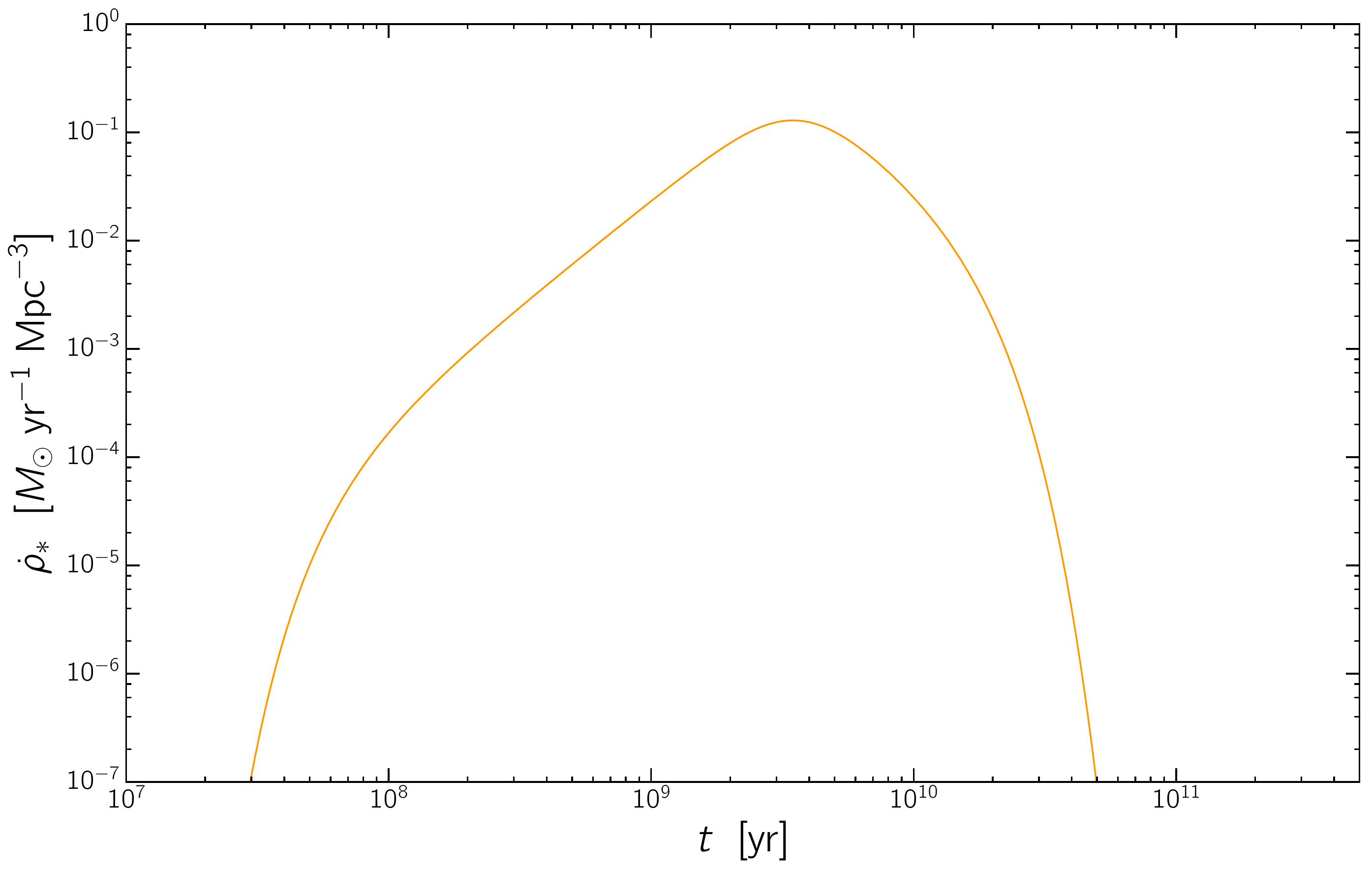}
	\includegraphics[width=.49\textwidth]{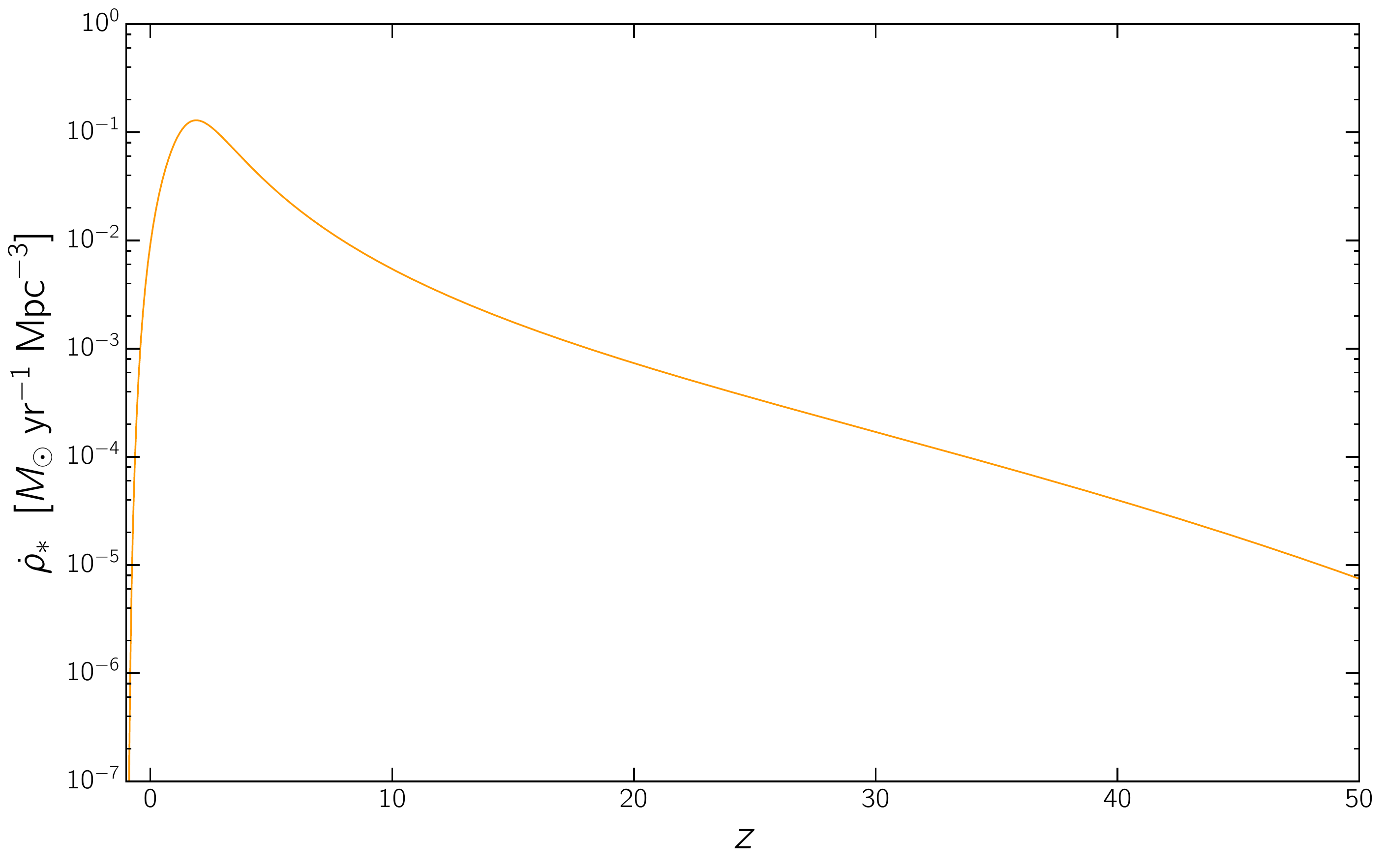}
	\caption{Star formation rate, $\dot{\rho}_*$, as a function of
          the cosmic time $t$ (left panel) and redshift $z$ (right
          panel), based on an extrapolation of a fitting function to
          existing data~\cite{MadauDickinson}.}
	\label{fig:sfr}
\end{figure}

The resulting star formation rate as a
function of time and redshift is shown in Figure~\ref{fig:sfr}.

\subsection{Probability of Life on a Habitable Planet}

The probability for the existence of life around a star of a
particular mass $m$ can be expressed in terms of the product between
the probability that there is an Earth-mass planet in the star's
habitable zone (HZ) and the probability that life emerges on such a
planet: $P({\rm life}| m)=P({\rm HZ} | m)P({\rm life | HZ}) $. The
first factor, $P({\rm HZ} | m)$, is commonly labeled $\eta_{\rm
  Earth}$ in the exo-planet literature~\cite{Traub}.

Data from the NASA Kepler mission implies $\eta_{\rm Earth}$ values in
the range of $6.4^{+3.4}_{-1.1}\%$ for stars of approximately a solar
mass ~\cite{EtaEarthSilburt,Petigura,Marcy} and $24^{+18}_{-8}\%$ for
lower mass M-dwarf stars~\cite{EtaEarthDC}.  The result for solar mass
stars is less robust due to lack of identified Earth-like planets at
high stellar masses. Owing to the large measurement uncertainties, we
assume a constant $\eta_{\rm Earth}$ within the range of stellar
masses under consideration. The specific constant value of $\eta_{\rm
  Earth}$ drops out of the calculation due to the normalization factor
$N$.

\begin{figure}[!h]
  \centering 
  \includegraphics[width=.94\textwidth]{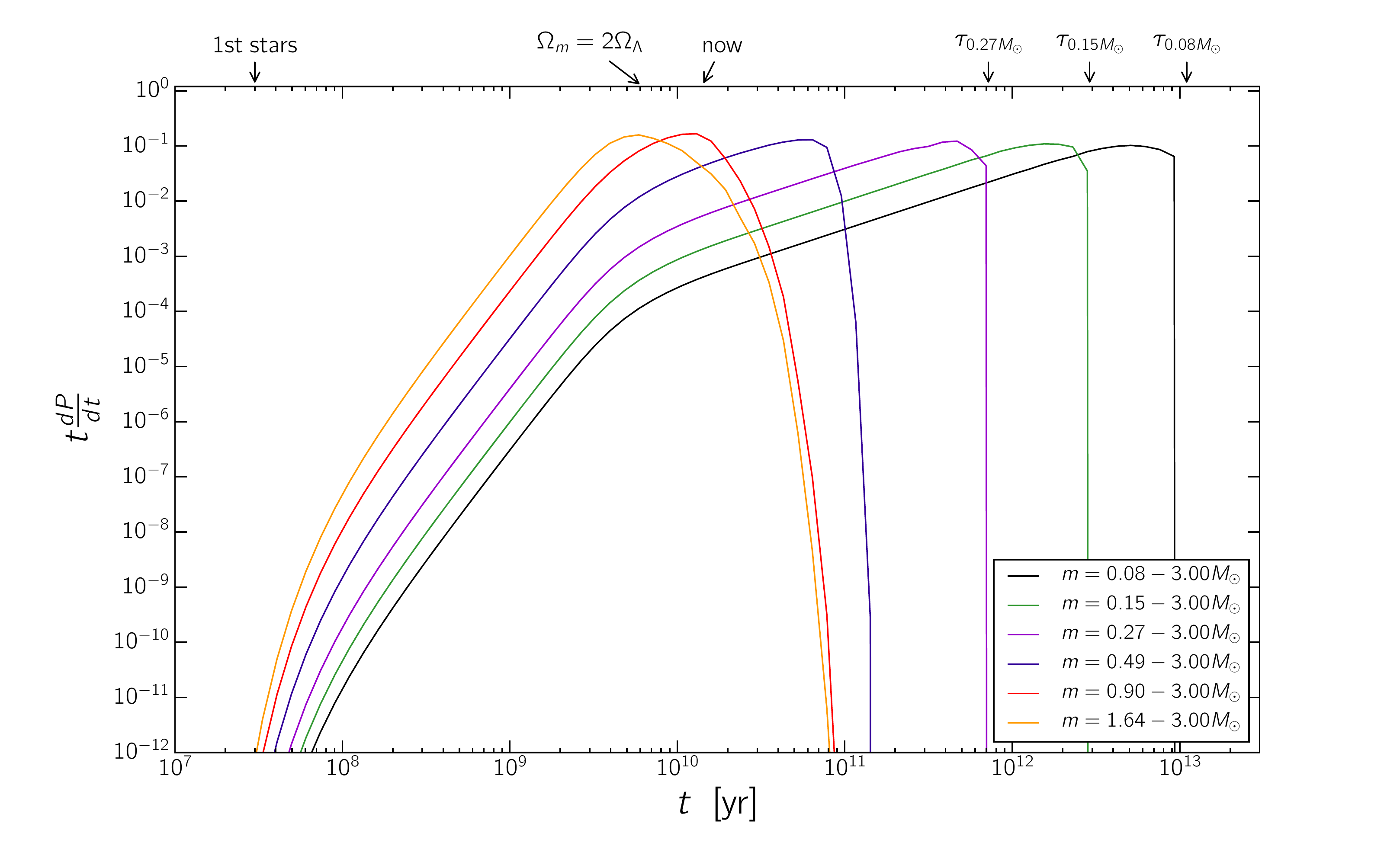} 
  \includegraphics[width=.94\textwidth]{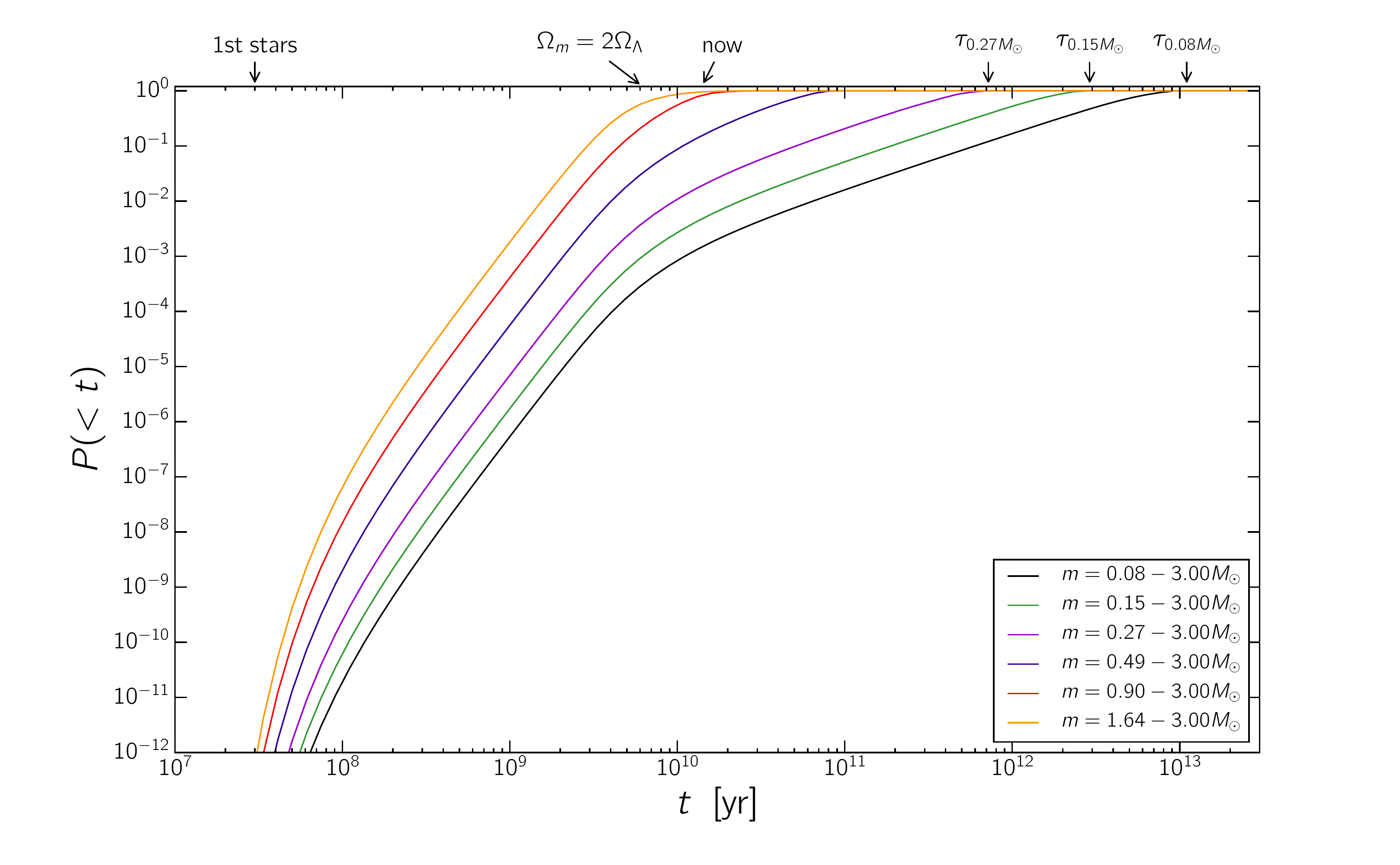} 
  \caption{Probability distribution for the emergence of life within a
  fixed comoving volme of the Universe as a function of cosmic
  time. We show the probability per log time, $tdP/dt$ (top panel) as
  well as the cumulative probability up to a time $t$, $P(<t)$ (bottom
  panel), for different choices of the minimum stellar mass, equally
  spaced in $\log m$ between $0.08M_\odot$ and $3M_\odot$. The
  contribution of stars above $3M_\odot$ to $dP(t)/dt$ is ignored due
  to their short lifetimes and low abundances. The labels on the top
  axis indicate the formation time of the first stars, the time when
  the cosmic expansion started accelerating (i.e., when the density
  parameter of matter, $\Omega_m$, was twice that of the vacuum,
  $\Omega_\Lambda$), the present time (now) and the lifetimes of stars
  with masses of $0.08M_\odot, 0.15M_\odot$ and $0.27M_\odot$.}  \label{fig:sft}
\end{figure}

There is scope for considerable refinement in the choice of the second
factor $p(\text{life}|\text{HZ})$. One could suppose that the
probability of life evolving on a planet increases with the amount of
time that the planet exists, or that increasing the surface area of
the planet should increase the likelihood of life beginning.  However,
given our ignorance we will set this probability factor to a constant,
an assumption which can be improved upon by statistical data from
future searches for biosignatures in the molecular composition of the
atmospheres of habitable
planets~\cite{Seager,LoebMaoz,Mercedes,LinLoeb}. In our simplified
treatment, this constant value has no effect on $dP(t)/dt$ since its
contribution is also cancelled by the normalization factor $N$.

\section{Results}
\label{Sec:3}

The top and bottom panels in Figure~\ref{fig:sft} show the probability
per log time interval $tdP(t)/dt=dP/d \ln t$ and the cumulative
probability $P(<t)=\int_0^t [dP(t')/dt'] dt'$ based on
equation~(\ref{eq:prob}), for different choices of the low mass cutoff
in the distribution of host stars for life-hosting planets equally
spaced in $\ln m$. The upper stellar mass cutoff has a negligible
influence on $dP/d \ln t$, due to the short lifetime and low abundance
of massive stars. In general, $dP/d\ln t$ cuts off roughly at the
lifetime of the longest lived stars in each case, as indicated by the
upper axis labels. For the full range of hydrogen-burning stars,
$dP(t)/d \ln t$ peaks around the lifetime of the lowest mass stars $t\sim
10^{13}~{\rm yr}$ with a probability value that is a thousand times
larger than for the Sun, implying that life around low mass stars in
the distant future is much more likely than terrestrial life around
the Sun today.

For mass ranges centered narrowly around a solar mass,
  we find that the current time indeed represents the peak of the probability
  distribution. However, as we allow lower mass stars to enter the
  distribution, the peak is shifted into the future due to the
  higher abundance of low mass stars and their extended
  lifetimes. It is interesting to note that a rather small extension
  of the mass range down to a quarter of a solar mass shifts the peak
  by two orders of magnitude in time into the future. Evidence of life around
  the nearest star, Proxima Centauri ($m=0.12 M_\odot$), for example, would indicate
  that our existence at or before the current cosmic time would be at
  the 0.1\% level.

\section{Discussion}
\label{Sec:4}

Figure~\ref{fig:sft} implies that the probability for life per
logarithm interval of cosmic time, $dP(t)/d\ln t$, has a broad distribution in
$\ln t$ and is peaked in the future, as long as life is likely around
low-mass stars. High mass stars are shorter lived and less abundant
and hence make a relatively small contribution to the probability
distribution.

Future searches for molecular biosignatures (such as O$_2$ combined with
CH$_4$) in the atmospheres of planets around low mass
stars~\cite{Seager,Mercedes} could inform us whether life will exist
at late cosmic times~\cite{Koppa}. If we insist that life near the Sun
is typical and not premature, i.e. require that the peak in
$dP(t)/d\ln t$ would coincide with the lifetime of Sun-like stars at
the present time, then we must conclude that the physical environments
of low-mass stars are hazardous to life (see
e.g. Ref.~\cite{raymond2007a}).  This could result, for example, from
the enhanced UV emission and flaring activity of young low-mass stars,
which is capable of stripping rocky planets of their
atmospheres~\cite{Owen}.

Values of the cosmological constant below the observed one should not
affect the probability distribution, as they would introduce only mild
changes to the star formation history due to the modified formation
history of galaxies~\cite{Nagamine,Busha}. However, much larger values
of the cosmological constant would suppress galaxy formation and
reduce the total number of stars per comoving volume~\cite{Loeb06},
hence limiting the overall likelihood for life
altogether~\cite{Weinberg}. This is of course a crude
  estimate, as we are varying only one fundamental parameter out of a
  plethora of possible changes in cosmological parameters. A larger
  amplitude of density fluctuations could certainly allow for
  structure formation to continue; however, Ref. ~\cite{tegmark1998a},
  for example, argues that this is anthropically disfavoured as it
  would lead to increased instability in planetary orbits and the
  habitable zones would not be occupied long enough for life to
  emerge.

Our results provide a new perspective on the so-called ``coincidence
problem'', {\it why do we observe $\Omega_m \sim
  \Omega_\Lambda$?}~\cite{carter1983a} The answer comes naturally if
we consider the history of Sun-like star formation, as the number of
habitable planets peaks around present time for $m \sim 1M_\odot$. We
note that for the majority of stars, this coincidence will not exist
as $dP(t)/dt$ peaks in the future where $\Omega_m \ll
\Omega_\Lambda$ (cf. Ref.~\cite{livio1999a,garriga2000a}). 
The question is then, why do we find ourselves
orbiting a star like the Sun now rather than a lower mass star in the
future?

One can certainly contend that our result presumes our
  existence, and we therefore have to exist at some time. Although our
  result puts the probability of finding ourselves at the current
  cosmic time within the 0.1\% level, rare events do happen. In this
  context, we reiterate that our results are an order of magnitude
  estimate based on the most conservative set of assumptions within
  the standard $\Lambda$CDM model. If one were to take into account
  more refined models of the beginning of life and observers, this
  would likely push the peak even farther into the future, and make
  our current time less probable. As an example, one could consider
  that the beginning of life on a planet would not happen immediately
  after the planet becomes `habitable'. Since we do not know the
  circumstances that led to life on Earth, it would be more
  realistic to assume that some random event must have occurred to
  initiate life, corresponding to a Poisson process. This would
  suppress early emergence and thus shift the peak probability to the
  future.

A similar effect would arise from refining our notion of a probability
through re-weighting by the number observers on a planet.  Although
this may reach some theoretical peak, populations are at best
exponentially suppressed to the past, and so a given {\it observer}
(as opposed to inhabited planet) would be more likely to exist later
on in a planet's existence. There are several other such factors
(stability of habitable zone, galactic mergers and reseeding, etc) all
of which push the peak to the future. Our conclusion is therefore that
the most conservative estimate put the probability of our existence
before the current cosmic time at 0.1\% at most. It is likely that
with more refined models this number will be reduced, and so the above
mentioned question becomes ever more pertinent.

We derived our numerical results based on a conservative set of
assumptions and guided by the latest empirical data for the various
components of equation (\ref{eq:prob}). However, the emergence of life
may be sensitive to additional factors that were not included in our
formulation, such as the existence of a moon to stabilize the climate
on an Earth-like planet~\cite{Laskar}, the existence of asteroid
belts~\cite{martin2013a}, the orbital structure of the host planetary
system (e.g. the existence of nearby giant planets or orbital
eccentricity), the effects of a binary star companion~\cite{Haghi},
the location of the planetary system within the host
galaxy~\cite{Linew}, and the detailed properties of the host galaxy
(e.g. galaxy type~\cite{Conroy} or metallicity~\cite{Johnson}),
including the environmental effects of quasars, $\gamma$-ray
bursts~\cite{Piran} or the hot gas in clusters of galaxies. These
additional factors are highly uncertain and complicated to model and
were ignored for simplicity in our analysis.

The probability distribution $dP(t)/d \ln t$ is of particular
importance for studies attempting to gauge the level of fine-tuning
required for the cosmological or fundamental physics parameters that
would allow life to emerge in our Universe.

\acknowledgments
AL thanks the ``Consolidation of Fine-Tuning'' project for inspiration
that led to this work. RAB and DS thank Harvard's Institute for
Theory \& Computation for its kind hospitality during the work on this
paper. RAB and DS acknowledge the financial support of the John Templeton 
Foundation.

\bibliographystyle{JHEP}
\bibliography{references}

\end{document}